\documentclass[12pt]{article}
\begin{document}
\title{{\bf BLACK HOLES WITH LESS
\\ ENTROPY THAN A/4}
\thanks{Alberta-Thy-14-01, gr-qc/0109040}}
\author{
Don N. Page
\thanks{Internet address:
don@phys.ualberta.ca}
\\
CIAR Cosmology Program, Institute for Theoretical Physics\\
Department of Physics, University of Alberta\\
Edmonton, Alberta, Canada T6G 2J1
}
\date{(2001 Sept. 11, revised 2001 Sept. 27)}
\maketitle
\large

\begin{abstract}
\baselineskip 16 pt

	One can increase one-quarter the area
of a black hole, $A/4$,
to exceed the total thermodynamic entropy, $S$,
by surrounding the hole
with a perfectly reflecting shell
and adiabatically squeezing it inward.
$A/4$ can be made to exceed $S$
by a factor of order unity
before the shell enters the Planck regime,
though practical limitations are much more restrictive.
One interpretation is that the black hole
entropy resides in its thermal atmosphere,
and the shell restricts the atmosphere
so that its entropy is less than $A/4$.

\end{abstract}
\normalsize
\baselineskip 16 pt
\newpage

\section{Introduction}

	The Generalized Second Law (GSL)
\cite{GSL}
of black hole thermodynamics states that
the total thermodynamic entropy $S$ does not decrease,
and it further states that for Einsteinian gravity
(to which this paper will be restricted,
though the generalization to various
other theories should be straightforward),
$S$ is the GSL entropy
 \begin{equation}
 S_{\rm GSL} \equiv {1\over 4}A + S_{\rm m},
 \label{eq:1}
 \end{equation}
where $A$ is the total event horizon area
of all black holes
and $S_{\rm m}$ is the entropy of matter
outside the black holes.
(I am using Planck units in which $\hbar$,
$c$, $4\pi\epsilon_0$, Boltzmann's constant $k$,
and the renormalized Newtonian gravitational constant $G$
are all set equal to unity.)
Although the Generalized Second Law
has only been proved under restricted conditions,
such as for quasistationary semiclassical
black holes
\cite{FP},
it is believed to have greater generality,
such as to rapidly evolving black holes.

	An implicit further assumption
that is often made is that
the matter entropy $S_{\rm m}$
cannot be negative.
This assumption, plus the GSL,
leads to the conclusion that the total
entropy is bounded below by one-quarter
the total event horizon area:
 \begin{equation}
 {1\over 4}A \leq S.
 \label{eq:3}
 \end{equation}

	Here I shall show that the inequality
(\ref{eq:3}) can be violated.
This violation can be interpreted as either
a violation of the Generalized Second Law
(if $S_{\rm m}$ is assumed to be restricted
to nonnegative values)
or as an indication that the matter entropy
$S_{\rm m}$ must be allowed to take negative values
in order to conform to the GSL.

	Briefly, a violation of
the inequality (\ref{eq:3}) can be produced
as follows:
Take a Schwarzschild black hole
of initial mass $M_i$ and radius $2M_i \gg 1$
(in the Planck units used herein)
with negligible matter outside
the hole and its nearby thermal atmosphere
(which is here taken to be part of the hole's
energy and entropy).
Assuming the GSL for this initial state,
the initial entropy $S_i$ is roughly
$A_i/4 = 4\pi M_i^2 \gg 1$,
one-quarter the initial area of the hole,
since the initial matter entropy $S_{\rm m}$
is negligible in comparison.
(If one considers as matter the thermal atmosphere
that forms when the horizon forms
in the near-horizon region $r-2M \ll 2M$,
either the entropy of this atmosphere
should be considered negligible
if it is considered to be part of $S_{\rm m}$,
or it should be considered to be part
of the black hole entropy $A/4$;
one gets too large a value for the total
entropy if one counts both $A/4$
and a large entropy associated with
the near-horizon thermal atmosphere.
There may indeed be a correction
to the entropy of the order of $\ln{A}$
from the thermal atmosphere and/or
from other considerations,
and this correction might be a few orders
of magnitude larger than unity for huge
black holes, but for the purposes of this paper
I shall not count such a correction as
``large''---here ``large'' numbers will mean those
many orders of magnitude larger than unity,
such as a positive power of the area.)

	Now surround the black hole by
a spherical perfectly reflecting shell
at a radius $r_i$ that is a few times
the Schwarzschild radius $2M_i$ of the
black hole.  This region
(outside of the near-horizon atmosphere
that is being counted as part of the hole)
will soon fill
up with thermal Hawking radiation
to reach an equilibrium state
of fixed energy $M_i$ inside the shell,
but for $M_i \gg 1$, all but a negligible
fraction ($\sim 1/M_i^2$ in Planck units)
of the energy will remain in the hole,
which can thus be taken still to have mass $M_i$.
Outside the shell, one will have essentially
the Boulware vacuum state with zero entropy
(plus whatever apparatus that one
will use to squeeze the shell in the next step,
but this will all be assumed to be in a pure state
with zero entropy).

	Next, squeeze the shell inward.
If this is done sufficiently slowly,
this should be an adiabatic process,
keeping the total entropy fixed.
Also, the outside itself should remain
in a zero-entropy pure state,
since the perfectly reflecting shell
isolates the region outside
from the region inside
with its black hole and thermal radiation,
except for the effects of the gravitational
field, which will be assumed to produce
negligible quantum correlations between
the inside and the outside of the shell
(as one would indeed get in a semiclassical
approximation in which the geometry
is given by a spherically symmetric classical metric).
Some of the thermal Hawking
radiation will thus be forced into the black hole,
increasing its area.

	So long as the shell is not taken
into the near-horizon region $r-2M \ll 2M$,
the radiation forced into the black hole
will have negligible energy and so will
not increase the black hole area significantly
above its initial value $A_i$.
(Indeed, some of this tiny increase in the area
just compensates for the tiny decrease
in the black hole area when it filled
the region $r < r_i$ with thermal radiation.)

	However, nothing in principle prevents one
from squeezing the shell into the near-horizon region,
where a significant amount of the near-horizon
thermal radiation can be forced into the hole,
increasing its mass $M$ and area $A = 4\pi M^2$
significantly.  Since the entropy $S$ should not change
by this adiabatic process, it remains very nearly at
$A_i/4$.  Therefore, one ends up with a squeezed
black hole configuration with $A > 4S \approx A_i$,
or total entropy significantly less than $A/4$.
(By significantly less, I mean that $A/4 - S$
is very large in absolute value,
not necessarily that it is a significant
fraction of $A/4$.)

	A simple way to interpret this result
is to say that the near-horizon thermal atmosphere
contributes a significant fraction (perhaps all)
of the black hole entropy.  Then when this atmosphere
is restricted to a smaller region by a near-horizon shell,
its contribution to the total entropy is reduced.

	Perhaps the simplest way to incorporate these
$S < A/4$ configurations into black hole thermodynamics
is modify the Generalized Second Law to state that
 \begin{equation}
 \tilde{S}_{\rm GSL} \equiv S_{\rm bh} + S_{\rm m}
 \label{eq:4}
 \end{equation}
does not decrease for a suitably coarse-grained
nonnegative $S_{\rm m}$ and for a suitable definition
of $S_{\rm bh}$ that reduces to $A/4 + O(\ln{A})$
(in Einstein gravity)
when there are no constraints on the near-horizon
thermal atmosphere but which is less than $A/4 + O(\ln{A})$
when the atmosphere is constrained
(and thus has less entropy).
One might interpret $S_{\rm bh}$ as arising
entirely from the near-horizon thermal atmosphere,
so that if the atmosphere is unconstrained in the vertical
direction, its entropy is at least approximately $A/4$.
(There is no fundamental difficulty in allowing that
in this unconstrained case, $S_{\rm bh}$ might also have other
smaller correction terms, such as a logarithm of the number
of fields or a logarithm of $A$ or of some other black hole
parameter.  It is just that in the unconstrained case,
the leading term of $S_{\rm bh}$ should be proportional to $A$,
and the coefficient should be $1/4$, at least in Einstein gravity.
I also do assume that in the unconstrained case
there is no other term in the black hole entropy
going as a positive power of $A$, such as $M = [A/(4\pi)]^{1/2}$.)
But if the near-horizon thermal atmosphere is
significantly constrained, it has much less entropy.

	An alternative (but perhaps less attractive)
way to incorporate these
$S < A/4$ configurations is to retain the
Generalized Second Law in the original form of Eq. (\ref{eq:1}),
which is the special case of Eq. (\ref{eq:4})
in which $S_{\rm bh} = A/4$,
but now to allow $S_{\rm m}$ to become negative
when one squeezes the black hole.
For example, one might use Eq. (\ref{eq:1})
not to define $S_{\rm GSL}$ in terms of $A/4$ and $S_{\rm m}$,
but instead to define $S_{\rm m}$ as the total
entropy $S_{\rm GSL}$ minus the black hole entropy $A/4$.
(Of course, this procedure would make the GSL useless for
telling what the total entropy is, so then $S_{\rm GSL}$
would have to be found by some other procedure.)

	A longer version of most of this paper has already appeared
\cite{BHentropy},
which the reader might like to consult for some details omitted here,
but the arguments are sharpened up and expressed more succinctly here.
One investigation that was pursued there, but not here,
is a closed-form approximation to the
static spherically symmetric metric obtained
by making a self-consistent nonlinear semiclassical gravitational
backreaction calculation with the expectation value
of the stress-energy tensor of the vacuum state outside a shell.
In contrast, here I shall confine myself to cases
in which the gravitational backreaction is sufficiently small
that it may be treated as a linear perturbation
to the Schwarzschild geometry.

\section{Calculation of the Entropy of a Black Hole \\
Inside a Perfectly Reflecting Shell}

	Let us now try to estimate
what the total entropy is of a configuration of
an uncharged, nonrotating black hole
of mass $M$ and area $A = 16\pi M^2$,
in equilibrium with Hawking thermal radiation inside
a perfectly reflecting pure-state shell
of radius R and local mass $\mu$,
outside of which one has vacuum.
This calculation is somewhat complicated,
as it involves specific assumptions about
the stress-energy tensor inside and outside
the shell and how the adiabatic motion of the shell
affects these.
In principle, these assumptions could be checked by
doing suitable calculations of quantum field theory
in nearly-static spacetimes with slowly-moving
perfectly reflecting boundaries, but these calculations
appear to be so difficult that I have replaced them
by what I believe are highly plausible
simple physical arguments.
Therefore, I do not have a rigorous proof of
the validity of my calculations,
but I think they are correct, and they do lead
to the approximate entropy formula (\ref{eq:49})
that seems to be eminently reasonable.

	As the beginning of the next section indicates,
my entropy formula can also easily be derived from
an even simpler set of assumptions that
are also plausible, though perhaps more open to question
than the ones I use in the derivation immediately below.
For the reader who accepts the validity of the simpler
assumptions of the next section, he or she may wish to
skip the present derivation and go immediately
to the results of Eqs. (\ref{eq:48}) and (\ref{eq:49})
at the end of this section.
However, anyone who has doubts about those assumptions
may find the present arguments and derivation instructive,
as they seem to me stronger than the simpler ones
of the next section. 

	We shall take a semiclassical approximation
with a certain set of matter fields,
which for simplicity will all be assumed to be massless
free conformally coupled fields.
Given the field content of the theory,
the three parameters $(M,R,\mu)$ determine the configuration,
though the entropy should depend only on $M$ and $R$,
since the shell and the vacuum outside have zero entropy.

	It is convenient to replace the shell radius $R$
with the classically dimensionless parameter
 \begin{equation}
 W = {2M \over R},
 \label{eq:5}
 \end{equation}
which would be 0 if the shell were at infinite radius
(though before one reached this limit the black hole
inside the shell would become unstable to evaporating away)
and 1 if the shell were at the black hole horizon
(though in this limit the forces on the shell
would have to be infinite).
Then we would like to find $S(M,W)$.

	If $W$ is neither too close to 0 nor to 1,
the entropy will be dominated by $A/4 = 4\pi M^2$.
The dominant relative correction to this will come
from effects of the thermal radiation and vacuum
polarization around the hole and so would have
a factor of $\hbar$ if I were using gravitational
units $(c=G=1)$ instead of Planck units $(\hbar=c=G=1)$.
In gravitational units,
$\hbar$ is the square of the Planck mass,
so to get a dimensionless quantity from that,
one must divide by $M^2$
(or by $R^2$, which is just $4 M^2/W^2$
with $W$ being of order unity);
for the free massless fields under consideration,
there are no other mass scales in the problem
other than the Planck mass.
Therefore, in Planck units,
the first relative correction to $A/4$
will have a factor of $1/M^2$
and hence give an additive correction term
to $4\pi M^2$ that is of the zeroth power of $M$.
Such a term could be a function of $W$,
and it could also involve the logarithm of
the black hole mass in Planck units,
since such a logarithm may be regarded as
being of the zeroth power of $M$.
However, it will not be proportional to
any positive power of $M$.

	One might expect that if one proceeded
further in this way, one would find
that the entropy $S$ is given by $4\pi M^2$
times a whole power series in $1/M^2$,
with each term but the zeroth-order one
having a coefficient that is a function of $W$
and of the logarithm of the black hole mass.
If we had been considering the possibility
of massive fields, then
these coefficients of the various powers
of $1/M^2$ would not be purely functions of $W$
but would also be functions of the masses of the fields.
However, for simplicity we shall consider only
the free massless field case here.
For simplicity I shall also mostly ignore the possible
dependence on the logarithm of the black hole mass,
so that the coefficients of the powers of $M$
will be assumed to be functions purely of $W$.

	In fact, I shall consider only the first
two terms in this power series and,
for simplicity, drop the possible $\ln{M}$ dependences:
 \begin{eqnarray}
 S(M,W) &=& 4\pi M^2 + f_1(W,\ln{M}) + f_2(W,\ln{M}) M^{-2} + \cdots
 \nonumber \\
 &\approx& 4\pi M^2 + f_1(W).
 \label{eq:6}
 \end{eqnarray}
The function $f_1(W)$ will depend on the
massless matter fields present in the theory,
most predominantly through the radiation constant
 \begin{equation}
 a_r = {\pi^2 \over 30}(n_b + {7 \over 8} n_f),
 \label{eq:7}
 \end{equation}
where $n_b$ is the number of bosonic helicity states
and $n_f$ is the number of fermionic helicity states
for each momentum.
It also proves convenient to define
 \begin{equation}
 \alpha \equiv {a_r \over 384\pi^3}
  = {n_b + {7 \over 8} n_f \over 11\,520\pi},
 \label{eq:8}
 \end{equation}
which makes the entropy density of the thermal Hawking radiation
far from the hole (when $R \gg M$ or $W \ll 1$) simply $\alpha/M^3$,
and also to set
 \begin{equation}
 f_1(W) = -32\pi\alpha s(W),
 \label{eq:9}
 \end{equation}
where $s(W)$ depends (weakly) only on the ratios of the numbers
of particles of different spins and so stays fixed
if one doubles the number of each kind of species.
Then my truncated power series expression for
the entropy of an uncharged spherical black hole of
mass $M$ at the horizon (and hence horizon radius $2M$
and horizon area $4\pi M^2$) surrounded by a
perfectly reflecting shell of radius $R = 2M/W$ is
 \begin{equation}
 S(M,W) \approx 4\pi M^2 - 32\pi\alpha s(W)
  = {1\over 4} A - 32\pi\alpha s(W).
 \label{eq:9b}
 \end{equation}

	Now I shall evaluate an approximate expression
for $s(W)$ when the perturbation to the Schwarzschild
geometry is small from the thermal radiation inside the shell
and from the vacuum polarization inside and outside the shell.
There will be an additive constant to $s(W)$
(possibly depending on $\ln{M}$), giving
an additive constant to the entropy, that I shall not
be able to evaluate, but for simplicity and concreteness
I shall assume that $s(1/2)=0$,
so that the entropy is $A/4$ when the shell is at $W=1/2$ or $R=4M$.

	First, I shall ignore the Casimir energy and related effects
of the shell itself on the fields.  I would expect that these
effects would give additive corrections to $s(W)$ that are of order
$W$ or smaller (and so never large compared with unity),
whereas the leading term in the perturbative approximation
for $s(W)$ will go as $1/W^3$
(proportional to the volume inside the shell) for $W \ll 1$
(shell radius $R \gg 2M$) and as $1/(1-W)$
(inversely proportional to the square of the redshift factor
to infinity) for $1-W \ll 1$
(shell radius relatively near the horizon),
so one or other of these leading terms
will dominate when $W$ is near 0 or 1.
Therefore, I shall take the stress-energy tensor inside the shell
to be approximately that of the Hartle-Hawking state
in the Schwarzschild geometry,
and that outside the shell to be approximately that
of the Boulware vacuum.

	The first part of the analysis will be done in
a coordinate system $(x^0,x^1,x^2,x^3) = (t,r,\theta,\phi)$
in which the spherically symmetric classical metric
has, at each stage of the process, the approximately static form
 \begin{equation}
 ds^2 = - e^{2\phi} dt^2 + U^{-1} dr^2
 	+ r^2 (d\theta^2 + \sin^2\theta d\varphi^2)
 \label{eq:10}
 \end{equation}
with
 \begin{equation}
 e^{2\phi} = e^{2\psi} U
 \label{eq:11}
 \end{equation}
and
 \begin{equation}
 U = 1 - {2m\over r} = 1-w
 \label{eq:12}
 \end{equation}
with
 \begin{equation}
 w \equiv {2m\over r} = 1-U = 1-(\nabla r)^2.
 \label{eq:13}
 \end{equation}
Here $\phi$, $U$, $\psi$, $m$, and $w$
are all functions of the $x^1=r$ coordinate alone,
although they also have a global dependence on
the black hole mass $M$ (the value of $r/2$
where $e^{2\phi}=0$), and (for $r>R$)
on the radius $r=R=2M/W$ of the shell and
on the total rest mass energy $\mu$ of the shell.
The Einstein equations then give
 \begin{equation}
 {d\psi\over dr} = 4\pi r (\rho + P) U^{-1}
 \label{eq:14}
 \end{equation}
and
 \begin{equation}
 {dm\over dr} = 4\pi r^2 \rho,
 \label{eq:15}
 \end{equation}
where
 \begin{equation}
 \rho = - \langle T^0_0(r)\rangle 
 \label{eq:16}
 \end{equation}
is the expectation value of the energy density
in the appropriate quantum state, and
 \begin{equation}
 P = \langle T^1_1(r)\rangle 
 \label{eq:17}
 \end{equation}
is the corresponding expectation value of the
radial pressure, both functions of $r$.
The functional form of the expectation value
of the tangential pressure
$\langle T^2_2(r)\rangle = \langle T^3_3(r)\rangle$
would then follow from the conservation
of $\langle T^{\mu}_{\nu}\rangle$ but will not be explicitly
needed in this paper.

	Since we are assuming that the
state of the quantum fields inside the shell
($r < R$) is the Hartle-Hawking
\cite{HH}
thermal state, for $r < R$ we have
 \begin{equation}
 \rho = \rho_H(M,r) \equiv {3\alpha\over 32 \pi M^4}\,\varepsilon_H(w)
 \label{eq:18}
 \end{equation}
and
 \begin{equation}
 P = P_H(M,r) \equiv {\alpha\over 32\pi M^4}\,p_H(w),
 \label{eq:19}
 \end{equation}
where on the extreme right hand side of each
of these two equations I have factored out
the dependence on the black hole mass $M$
from that on the
radial function $w \equiv 2m/r$
that is classically dimensionless
(dimensionless without setting $\hbar = 1$),
thereby defining two classically dimensionless
functions of $w$, $\varepsilon_H(w)$ and $p_H(w)$.

	Similarly, we are assuming that
the state of the quantum fields outside the shell
($r > R$) is the Boulware
\cite{Boul}
vacuum state, so for $r > R$ we have
 \begin{equation}
 \rho = \rho_B(M_{\infty},r)
  \equiv {3\alpha\over 32 \pi M_{\infty}^4}\varepsilon_B(w)
 \label{eq:20}
 \end{equation}
and
 \begin{equation}
 P = P_B(M_{\infty},r) \equiv {\alpha\over 32\pi M_{\infty}^4}p_B(w),
 \label{eq:21}
 \end{equation}
thereby defining two new classically dimensionless
functions of $w$, $\varepsilon_B(w)$ and $p_B(w)$.
Here
 \begin{equation}
 M_{\infty} \equiv m(r=\infty)
 \label{eq:22}
 \end{equation}
is the ADM mass at radial infinity.

	In some cases one can assume that there
is some extra apparatus in the region $r > R$ holding
the shell in.  If so, its energy density and radial pressure
can simply be included in $\rho_B$ and $P_B$.
In any case, we shall assume that the shell,
and whatever is outside the shell,
is in a pure state with zero entropy.
Therefore, the only contribution to the entropy
will come from the interior to the shell.

	Below we shall also need the vacuum polarization
part of the stress-energy tensor inside the shell,
whose components I shall denote by
 \begin{eqnarray}
 &&\rho_V(M,r) \equiv \rho_H(M,r) - \rho_T(M,r)
 \nonumber \\
 &&\equiv {3\alpha\over 32 \pi M^4}\varepsilon_V(w)
 \equiv {3\alpha\over 32 \pi M^4}(\varepsilon_H(w) - \varepsilon_T(w))
 \label{eq:23}
 \end{eqnarray}
and
 \begin{eqnarray}
 &&P_V(M,r) \equiv P_H(M,r) - P_T(M,r)
 \nonumber \\
 &&\equiv {\alpha\over 32 \pi M^4}p_V(w)
 \equiv {\alpha\over 32 \pi M^4}(p_H(w) - p_T(w)),
 \label{eq:24}
 \end{eqnarray}
where $\rho_T$ and $P_T$ denote the components
of the thermal parts.

	I shall assume that the vacuum polarization part
is what the Boulware state would give if one had it
inside the shell, so that,
in my approximation of ignoring Casimir effects,
$\rho_V$ and $P_V$
have the same dependence on the local mass $m(r)$
and radius $r$ as $\rho_B$ and $P_B$ do outside
the shell (when there is no extra apparatus there).
In the first-order (in $\alpha/M^2$)
perturbative calculation being done here,
the expectation value of the stress tensor
is already first order (except possibly for that of the shell),
so its functional dependence
on $m$ can be replaced by its dependence on its
zeroth approximation, which is the black hole mass $M$ for $r < R$
and the ADM mass $M_{\infty}$ for $r > R$.
Therefore, to sufficient accuracy for our purposes,
$\rho_V$ and $P_V$ can be evaluated by using
Eqs. (\ref{eq:20}) and (\ref{eq:21}) for $\rho_B$ and $P_B$
with the ADM mass $M_{\infty}$, which is approximately
the value of the local mass $m(r)$ anywhere outside the massive shell,
replaced by the black hole mass $M$,
which is approximately the value of $m(r)$ anywhere inside the shell.
In particular, this implies that we can use
 \begin{equation}
 \varepsilon_V(w) = \varepsilon_B(w)
 \label{eq:25}
 \end{equation}
and
 \begin{equation}
 p_V(w) = p_B(w).
 \label{eq:26}
 \end{equation}

	For explicit approximate calculations,
it is useful to have explicit approximate formulas
for these various components of the stress-energy tensor
(though only some of these are necessary
for the final result to be given below),
given by the equations above from the six functions
$\varepsilon_H(w)$, $\varepsilon_B(w)$,
$\varepsilon_T(w) = \varepsilon_H(w) - \varepsilon_B(w)$,
$p_H(w)$, $p_B(w)$, and $p_T(w) = p_H(w) - p_B(w)$.
For simplicity and concreteness, I shall use those
obtained for a conformally invariant massless scalar field
in the gaussian approximation
\cite{Page82}, which gives
 \begin{eqnarray}
 \varepsilon_H(w) &\equiv& {32\pi M^4\over 3\alpha}\rho_H
 \equiv {(8\pi M)^4\over a_r}\rho_H
 \approx {1-(4-3w)^2 w^6\over (1-w)^2} - 24w^6
 \nonumber \\
 &=& 1+2w+3w^2+4w^3+5w^4+6w^5-33w^6,
 \label{eq:27}
 \end{eqnarray}
 \begin{eqnarray}
 \varepsilon_B(w) &\equiv& {32\pi M_{\infty}^4\over 3\alpha}\rho_B
 \equiv {(8\pi M_{\infty})^4\over a_r}\rho_B
 \approx {-(4-3w)^2 w^6\over (1-w)^2} - 24w^6
 \nonumber \\
 &=& -{1\over (1-w)^2}+1+2w+3w^2+4w^3+5w^4+6w^5-33w^6, 
 \label{eq:28}
 \end{eqnarray}
 \begin{equation}
 \varepsilon_T(w) = \varepsilon_H(w) - \varepsilon_B(w)
 \approx {1\over (1-w)^2} = {1\over U^2},
 \label{eq:29}
 \end{equation}
 \begin{eqnarray}
 p_H(w) &\equiv& {32\pi M^4\over \alpha}P_H
 \equiv {(8\pi M)^4\over 3 a_r}P_H
 \approx {1-(4-3w)^2 w^6\over (1-w)^2} + 24w^6
 \nonumber \\
 &=& 1+2w+3w^2+4w^3+5w^4+6w^5+15w^6,
 \label{eq:30}
 \end{eqnarray}
 \begin{eqnarray}
 p_B(w) &\equiv& {32\pi M_{\infty}^4\over \alpha} P_B
 \equiv {(8\pi M_{\infty})^4\over 3 a_r} P_B
 \approx {-(4-3w)^2 w^6\over (1-w)^2} + 24w^6
 \nonumber \\
 &=& -{1\over (1-w)^2}+1+2w+3w^2+4w^3+5w^4+6w^5+15w^6, 
 \label{eq:31}
 \end{eqnarray}
and
 \begin{equation}
 p_T(w) = p_H(w) - p_B(w)
 \approx {1\over (1-w)^2} = {1\over U^2}.
 \label{eq:32}
 \end{equation}

Note that this approximation gives
 \begin{equation}
 \rho_T \approx 3P_T \approx a_r T_{\rm local}^4,
 \label{eq:33}
 \end{equation}
just like thermal radiation in flat spacetime,
where $T_{\rm local}$ is the local value of
the Hawking temperature,
 \begin{equation}
 T_{\rm local} \approx {1 \over 8\pi m}(1-{2m\over r})^{-1/2}.
 \label{eq:34}
 \end{equation}
The form of $s(W)$ to be calculated actually
depends only on $\rho_T$ and $P_T$,
so any stress-energy tensor in which they
have the massless thermal form given above
will give the same results for the leading
correction to the entropy from the position of the shell
(when possible $\ln{M}$ terms are neglected).

	Now we use the Einstein equations
(\ref{eq:14}) and (\ref{eq:15})
with the appropriate $\rho$ and $P$
on the right hand side, and with the metric function $U$
there taking on its approximate Schwarzschild form,
$1-2M/r$ for $r < R$ and $1-2M_{\infty}/r$ for $r > R$.

	We also need to consider the effect of the shell,
which has a surface stress-energy tensor with components
 \begin{equation}
 S^0_0 = - {\mu\over 4\pi R^2}
 \label{eq:35}
 \end{equation}
and
 \begin{equation}
 S^2_2 = S^3_3 = - {F\over 2\pi R},
 \label{eq:36}
 \end{equation}
where $\mu$ is the total local mass of the shell,
the shell area $4\pi R^2$ multiplied by the
local mass-energy per area $-S^0_0$ as seen by a local
observer fixed on the shell,
and $F$ is the local total tensile force pulling
together the two hemispheres of the shell,
the circumference $2\pi R$ multiplied by
the local surface tension (tensile force per length)
$-S^2_2 = -S^3_3$.

	If one integrates the Einstein equations
(\ref{eq:14}) and (\ref{eq:15}) through the shell
and uses the conservation law for the stress-energy tensor,
one get the junction conditions
\cite{junc}
in the static case that
 \begin{equation}
 \mu = R (U_{-}^{1/2} - U_{+}^{1/2})
 \label{eq:37}
 \end{equation}
and
 \begin{equation}
 8F = {\mu\over R} + (1+8\pi R^2 P_{-})U_{-}^{-1/2}
  - (1+8\pi R^2 P_{+})U_{+}^{-1/2},
 \label{eq:38}
 \end{equation}
where
 \begin{equation}
 U_{-} = 1 - {2M_{-}\over R}
 \label{eq:39}
 \end{equation}
is the value of $U$ just inside the shell ($r = R-$),
where the local mass function $m$ takes on the value $M_{-}$,
and
 \begin{equation}
 U_{+} = 1 - {2M_{+}\over R}
 \label{eq:40}
 \end{equation}
is the value of $U$ just outside the shell ($r = R+$),
where the local mass function $m$ takes on the value $M_{+}$.
Similarly, $P_{-}$ and $P_{+}$ are the expectation values of the
radial pressure of the respective quantum states just inside
and just outside the shell.

	Thus we have at least five relevant masses
for the configuration:  the black hole mass $M = m(r=2M)$,
the mass $M_{-} = m(r=R-)$ just inside the shell,
the local mass (or local energy) $\mu$
of the shell itself at radius $r=R$,
the mass $M_{+} = m(r=R+)$ just outside the shell,
and the ADM mass $M_{\infty} = m(r=\infty)$ at radial infinity.
Since the stress-energy tensor inside the shell
is that of the Hartle-Hawking state determined by $M$ and $r$,
$M_{-}$ is a function of $M$ and $R$.
Similarly, since the stress-energy tensor outside the shell
is that of the Boulware state determined by $M_{\infty}$ and $R$
(at least when we do not have an extra apparatus there
to hold the shell in place),
$M_{+}$ is a function of $M_{\infty}$ and $R$.
The junction condition (\ref{eq:27}) then gives $\mu$
as a function of $M_{-}$, $M_{+}$, and $R$,
and hence as a function of $M$, $M_{\infty}$, and $R$.
One can in principle invert this to get $M_{\infty}$
(and hence each of the other masses as well)
as a function of $M$, $R$, and $\mu$,
or to get $M$ and each other mass as a function
of $M_{\infty}$, $R$, and $\mu$.
The main point is that if we just have a black hole
with the Hartle-Hawking thermal state inside a shell,
and the Boulware vacuum state outside the shell,
the semiclassical configuration (for fixed field
content of the quantum field theory)
is determined by three parameters,
though only two of them
(say $M$ and either $R$ or $W=2M/R$)
are relevant for the entropy
which resides purely inside the shell.

	To evaluate the function $s(W)$
in the truncated entropy formula (\ref{eq:9b}),
I shall consider an adiabatic process of slowly
squeezing the shell, keeping the total entropy constant
and thereby getting
 \begin{equation}
 {ds\over dW} = {M\over 4\alpha}{dM\over dW}
 \label{eq:41}
 \end{equation}
during this process.
Since this process is not strictly static,
one cannot use precisely the static metric (\ref{eq:1})
with $\phi$ and $U$ (or $\psi$ and $m$) that are
purely functions of $r$
and obey the static Einstein equations
(\ref{eq:14}) and (\ref{eq:15}).
However, one can consider a quasi-static metric
in which $\phi$ and $U$ (or $\psi$ and $m$)
have a very slow dependence on the time coordinate $t$
and the Einstein equations are only slightly different
from Eqs. (\ref{eq:14}) and (\ref{eq:15}).

	The specific calculation which I shall do
will be to have the shell squeeze itself inward by using
its own internal energy, so that no apparatus is used
outside the shell to push it inward,
and so that that outside region
has only the Boulware vacuum polarization.
The contraction of the shell is assumed to be so slow
that it does not excite the vacuum outside it
but rather leaves it in the Boulware vacuum state
with constant $M_{\infty}$.
However, as the shell moves in, it is enlarging
the Boulware state region, so effectively the shell
must be creating a larger volume of vacuum
with its vacuum polarization.
This means that in the slowly inmoving frame of the shell,
there is a flux of energy from the shell into the Boulware region,
needed to enlarge the Boulware region
while keeping it static where it already exists.
[For the stress-energy tensor components of
the Boulware vacuum given by Eqs. (\ref{eq:28})
and (\ref{eq:31}), this energy influx into
the Boulware region is actually negative,
so it increases the energy of the shell as it moves inward.]

	Similarly, if the inside of the shell
were also vacuum that did not get excited by
the adiabatic contraction of the shell,
there would be a swallowing up of part of the vacuum
region by the shell as it moves inward.
This would give a flux of (negative) energy
from the vacuum inside into the shell,
decreasing its energy.
Surely this flux into the shell also
exists even if the inside is not vacuum,
and I assume that it is given by the vacuum
polarization part of the actual stress-tensor there,
which I take to be approximately that of a Boulware state
with the same $m$ and $r$.
The remaining part of the total stress-energy tensor
there, which I am calling the thermal part,
and which is given approximately by Eq. (\ref{eq:33}),
should simply be reflected by the shell and
not give an energy flux into it
(in the frame of the slowly contracting shell),
though it will contribute to the force that needs
to be counterbalanced very nearly precisely
to obey the static junction equation (\ref{eq:38})
to high accuracy in order that the shell not have any
significant acceleration relative to a static frame.

	In other words, I am assuming that if a shell
moves inward through a static geometry, the vacuum
polarization part of the stress-energy tensor will stay static,
with $T_0^0 = - \rho_V(M,r)$, $T_1^1 = P_V(M,r)$,
and $T_0^1 = T_1^0 = 0$ inside the shell, and with
$T_0^0 = - \rho_B(M_{\infty},r)$, $T_1^1 = P_B(M_{\infty},r)$,
and $T_0^1 = T_1^0 = 0$ outside the shell.
Then as the shell moves through this static stress-energy
tensor, in the frame of the shell, there will
be fluxes of energy into or out from the shell
on its two sides.  In contrast, I am assuming that
the thermal radiation part of the stress-energy tensor
will be perfectly reflected by the shell,
so that in the frame of the shell it will give
no energy fluxes into or out from the shell.

	There is a modification of this picture that occurs
when the inward motion of the shell squeezes thermal radiation
into the black hole so that its mass goes up.
While the hole mass is increasing, the vacuum polarization
inside the shell is not quite static but instead has
small $T_0^1$ and $T_1^0$ terms that, for sufficiently
slow adiabatic processes, are proportional to $\dot{M}$,
the coordinate time derivative of the black hole mass $M$.
In the present calculation, in which the shell is squeezing itself
inward by using its own internal energy,
the ADM mass $M_{\infty}$ stays fixed, and so
the vacuum stress-energy tensor outside the shell
stays static during the process, under my approximation
of neglecting Casimir-type boundary effects of the
shell itself on the quantum field.
For a sufficiently slow inward squeezing of the shell,
the $T_0^1$ and $T_1^0$ terms inside are small, but over
the correspondingly long time of the squeezing they contribute
an effect on the energy balance of the shell that is not
completely negligible when one contemplates squeezing the shell
to a final position very near the black hole horizon.

	My procedure for calculating the small
$T_0^1$ and $T_1^0$ terms inside the shell is to assume that
the shell squeezing, and all consequent processes,
occur so slowly that $T_0^0$ and $T_1^1$
are given to high accuracy by the same functions
of $M$ and $r$ as they are when the geometry is static,
namely $-\rho_V(M,r)$ and $P_V(M,r)$.
Then I assume that the vacuum polarization part of
the stress-energy tensor is itself conserved
away from the shell,
so one can use the conservation of its energy
to deduce the radial derivative of $e^{\psi}r^2 T_0^1$.

	In particular, if we let the vacuum polarization
part of the stress-energy tensor have the component
 \begin{equation}
 T_0^1 = {\alpha\dot{M}w^2\over 4\pi M^4}e^{-\psi}f,
 \label{eq:42a}
 \end{equation}
with the factors chosen so that $f$ is a function purely of $w$,
then $T^{\mu}_{0;\mu}=0$ becomes
 \begin{equation}
 {\partial f\over\partial r} = {\pi M^2 e^{\psi}r^2\over\alpha\dot{M}}
 [\dot{\rho}_V + {\dot{m}\over rU}(\rho_V + P_V)].
 \label{eq:42b}
 \end{equation}
For the region inside the shell
with $r$ not too much larger than $2M$,
one has $m \approx M$ and $e^{\psi} \approx 1$
(possibly after suitably normalizing the time coordinate $t$).
Then if one uses Eqs. (\ref{eq:23})-(\ref{eq:26}),
one can rewrite Eq. (\ref{eq:42b}) as
 \begin{equation}
 {df\over dr} =
  -{3w\over 4}{d\over dw}\left({\varepsilon_B \over w^4}\right)
  -{3\varepsilon_B + p_B \over 8w^3(1-w)}.
 \label{eq:42c}
 \end{equation}
Given the functions $\varepsilon_B(w)$ and $p_B(w)$,
e.g., as given by Eqs. (\ref{eq:28}) and (\ref{eq:31})
from the gaussian approximation for a conformally
invariant massless scalar field, one can integrate Eq. (\ref{eq:42c})
to obtain $f(w)$ up to a constant of integration.
Although the constant of integration is not important,
it can also be fixed
by assuming that an observer that remains at fixed $w = 2m/r$
as $m$ changes sees in its frame no energy flux
in the limit that $w$ is taken to unity,
which implies that the flux of vacuum polarization energy
through the horizon is taken to be zero.

	After one calculates the vacuum polarization part
of the stress-energy tensor, which gives
$T_1^1-T_0^0 = \rho_B + P_B$ and $T_0^1 = 0$
outside the shell and
$T_1^1-T_0^0 = \rho_V + P_V$ and $T_0^1$
as given by Eq. (\ref{eq:42a}) inside the shell,
one can then calculate the fluxes of energy out from and
into the shell and insert these into the conservation
equations for the surface stress-energy tensor of the shell.
For a very slowly expanding or contracting shell, one finds that
 \begin{equation}
 d\mu = 4F dR
  + 4\pi R^2 dR[(\rho_B+P_B)U_{+}^{-1/2} - (\rho_V+P_V)U_{-}^{-1/2}]
  - 4\pi R^2 T_0^1 U_{-}^{-1/2} dt.
 \label{eq:42d}
 \end{equation}
The first term on the right hand side is the work done by
the tensile force within the shell,
and the remaining terms are the energy input from the
vacuum stress-tensor components $\rho_B$ and $P_B$
just outside the shell and the
vacuum stress-tensor components $\rho_V$, $P_V$, and $T_0^1$
just inside the shell.

	One now combines this local energy
conservation equation for the shell with the static junction
equations (\ref{eq:37}) and (\ref{eq:38})
that should still apply to high accuracy
in this slowly evolving situation to
keep the shell radius from accelerating
too rapidly.
When one also combines this with the
integrals of Eq. (\ref{eq:15}),
 \begin{equation}
 M_{-} = M + \int_{2M}^R 4\pi r^2 dr \rho_H,
 \label{eq:43}
 \end{equation}
 \begin{equation}
 M_{+} = M_{\infty} - \int_R^{\infty} 4\pi r^2 dr \rho_B,
 \label{eq:44}
 \end{equation}
one finds
 \begin{equation}
 \left(1 - {4\alpha f \over M^2}\right) dM
 \approx - 4\pi R^2 dR (\rho_T + P_T)
 \label{eq:45}
 \end{equation}
during the adiabatic contraction of the shell,
which, up to the small correction factor involving $f$,
is precisely what one would get
in flat spacetime from adiabatically
contracting a ball of thermal radiation.

	Next, we can use the fact that
$R = 2M/W$ to derive that
 \begin{equation}
 {dW \over dM} = {2\over R}\left(1-{M\over R}{dR\over dM}\right)
 \approx {2\over R}
 \left[1+{M(1-4\alpha f/M^2)\over 4\pi R^3(\rho_T + P_T)}\right],
 \label{eq:46}
 \end{equation}
where $f$ and $\rho_T + P_T$ are to be evaluated at $r=R$
or $w\approx W$.
Inserting this back into Eq. (\ref{eq:41})
then gives
 \begin{equation}
 {ds\over dW}
 \approx {3\varepsilon_B + p_B \over 4W^4}
  \left\{1+{4\alpha\over M^2}
   \left[{3\varepsilon_B + p_B\over 4W^3}-f\right]\right\}^{-1}.
 \label{eq:47}
 \end{equation}
For massless particles of any spin,
it should be a good approximation to take
$\rho_T + P_T \approx (4/3) a_r T_{\rm local}^4$ in terms of
the local temperature $T_{\rm local}$, and this implies that
$3\varepsilon_B + p_B \approx 4/(1-W)^2$, so
 \begin{equation}
 {ds\over dW}
 \approx {1 \over W^4 (1-W)^2}
  \left\{1+{4\alpha\over M^2}
   \left[{1 \over W^3 (1-W)^2}-f\right]\right\}^{-1}.
 \label{eq:47b}
 \end{equation}
 
	If we omitted the $f$ term from the radial
flux of vacuum polarization energy when $M$ changes,
then the factor inside the curly brackets above would
diverge as one approached the horizon, where $W=1$.
This implies that the reciprocal of this factor would cancel
the divergence in the factor before it, so $ds/dW$
would stay finite all the way down to $W=1$,
and one would find that the increase of one-quarter the area over
the entropy would be limited to an amount of the order
of $\sqrt\alpha M$.  For large $M$ this is large in absolute
units, but it is always much smaller than the entropy
itself, which is of the order of $4\pi M^2$.

	However, one can use the fact that
the regularity of the Hartle-Hawking stress-energy
tensor at the horizon implies that $\rho_H + P_H$,
and hence $3\varepsilon_H + p_H$, must go to zero
at least as fast as $1-w$ as one approaches the horizon.
(This is easiest to see in the Euclidean section
with imaginary time $t$,
on which for fixed coordinates $\theta$ and $\varphi$,
the horizon is at the center of a regular rotationally
symmetric two-surface with angular coordinate
$i\kappa t$ with $\kappa \approx 1/(4M)$ being the black hole
surface gravity and with the radial distance being
roughly $4M\sqrt{1-w}$ when $1-w \ll 1$.
Then $P_H = T_1^1$ is the pressure in the radial direction,
and $-\rho_H = T_0^0$ is the Euclidean pressure in the
Euclidean angular direction, and regularity at the origin
demands that the difference go to zero at least as fast
as the square of the radial distance from the origin.)
Then one can show that $f$ cancels the divergence
in $W^{-3}(1-W)^{-2}$ so that $W^{-3}(1-W)^{-2}-f$
stays finite as one approaches the horizon.
In fact, if one chooses the constant of integration of $f$
so that the flux of vacuum polarization energy through
the horizon is zero as $M$ is slowly changed,
then $W^{-3}(1-W)^{-2}-f$ actually goes to zero linearly
with $1-W$ as one approaches the horizon.
For example, using this constant of integration and
the gaussian approximation for $3\varepsilon_B$ and $p_B$
leads to
 \begin{equation}
 {ds\over dW}
 \approx {1 \over W^4 (1\!-W)^2}
  \left\{1\!+\!{4\alpha\over M^2 W^3}
   (1\!-\!W)(1\!+\!3W\!+\!6W^2\!+\!2W^3\!+\!7W^4\!+\!13W^5)\right\}^{-1}.
 \label{eq:47c}
 \end{equation}
 
 	Therefore, we see that the correction term
that is first order in $\alpha/M^2$ inside the curly brackets
of Eqs. (\ref{eq:47})-(\ref{eq:47c}) does not diverge
as one takes $1-W$ to zero but instead always remains small.
Therefore, we can drop it (as we have also neglected other
finite corrections that are linear in $\alpha/M^2$) and
integrate the zeroth-order part of Eq. (\ref{eq:47c})
to get an explicit formula for $s(W)$:
 \begin{equation}
 s(W) \approx \int_{1/2}^W {dw \over w^4(1-w)^2}
 = {1\over 1-W}+4\ln{W\over 1-W}-{1\over 3W^3}
 -{1\over W^2}-{3\over W}+{32\over 3}.
 \label{eq:48}
 \end{equation}
As discussed above, I arbitrarily chose the constant of integration
of this integral to make $s(W) = 0$ at $W = 1/2$ or $R = 4M$,
but this is not likely to be valid,
and there are also Casimir-energy effects from the
shell and corrections to Eqs. (\ref{eq:29}) and (\ref{eq:32})
that would give correction terms at least of order $W$
and likely also of the order of a constant and of order $1/W$.
From the logarithmic terms I am also ignoring,
I would also expect there to be corrections of the order
of $\ln{M}$.

	Finally, we can insert this form for $s(W)$
into Eq. (\ref{eq:9b}) to get
 \begin{eqnarray}
 S(M,W)\!\!\! &=& \!\!\!
  {1\over 4} A [1 - {8\alpha\over M^2} s(W)]
 = 4\pi M^2 -32\pi\alpha s(W)
 \nonumber \\
 &\approx& \!\!\!
  4\pi M^2 - 32\pi\alpha\left[{1\over 1-W}+4\ln{W\over 1-W}
        - {1\over 3W^3} - {1\over W^2}
	 + O\left({1\over W}\right)\right]
 \nonumber \\
 &\approx& \!\!\!
  4\pi M^2 - 32\pi\alpha\left[{R\over R-2M}+4\ln{2M\over R-2M}
        - {R^3\over 24M^3} - {R^2\over 4M^2}\right]
 \nonumber  \\
 &=& \!\!\! 4\pi M^2 \! - \! {n_b \! + \! {7 \over 8} n_f \over 360}
 	\left[{1\over 1 \! - \! 2M/R}
		\! - \! 4\ln{\left({R\over 2M} \! - \! 1\right)}
        	\! - \! {R^3\over 24M^3}
		 \! - \! {R^2\over 4M^2}\right],
 \label{eq:49}
 \end{eqnarray}
where after the second approximate equality
I have dropped the $O({1\over W})$ terms in Eq. (\ref{eq:48})
that I suspect are always dominated by corrections
to my approximations that I have not included.
Although I have retained four terms from $s(W)$
inside the square brackets,
only the first two terms should be kept when
$1-W = 1-2M/R \ll 1$ (shell very near the horizon),
and only the next two terms should be retained
when $W = 2M/R \ll 1$ (shell very large compared with the black hole).

	One might question the validity of getting
a large energy influx into the black hole horizon
from squeezing the reflecting shell
deep into the near-horizon region.
In my calculation it came from assuming that
the vacuum-polarization part of the stress-energy tensor
inside the contracting shell is absorbed by the shell
and hence has no effect on changing the mass of the hole,
whereas the thermal-radiation part is perfectly reflected
inward by the shell and hence increases the black hole mass.
But someone might object that the total stress-energy
tensor inside the shell is small, so that manipulating
it would not seem to be able to increase the black hole
mass significantly.

	However, another way of seeing that a significant
increase of the black hole mass is reasonable is to
consider the fact that the ADM mass at infinity,
$M_{\infty}$, is fixed,
and that as the shell is moved inward, it opens up
a larger and larger region of Boulware vacuum outside it,
which has negative energy density.
Therefore, for fixed $M_{\infty}$,
the mass just outside the shell, $M_{+}$,
increases as the shell is moved inward.
Because the inward-moving shell is
converting part of its local energy $\mu$
into doing work against the pressure difference across it
(greater pressure on the inside from
the Hawking radiation inside,
or one could equivalently say greater tension on the outside
from the fact that one has the Boulware state outside),
the mass just inside the shell, $M_{-}$,
increases even more than $M_{+}$ does
as the shell is moved inward.
The small total value of the energy density
inside the shell means that the black hole mass, $M$,
is very nearly the same as $M_{-}$
and hence increases significantly
as the reflecting shell is moved deep
into the near-horizon atmosphere of the black hole.

\section{Alternative Justifications of the Black Hole \\
Entropy Formula}

	The result indicated by Eqs. (\ref{eq:48})
and (\ref{eq:49})
is precisely the same that one would obtain by taking
the geometry to be Schwarzschild with a thermal bath
of radiation with local Hawking temperature
 \begin{equation}
 T_{\rm local} = {1 \over 8\pi M}(1-{2M\over r})^{-1/2}
 \label{eq:50}
 \end{equation}
and entropy density $(4/3)a_r T_{\rm local}^3$,
and then taking the total entropy to be $4\pi M^2$
plus the entropy difference between that inside the
shell at radius $R$ and that inside the radius $4M$.
If one na\"{\i}vely integrates this assumed entropy density
all the way down to the horizon, one would get a divergence,
but one can take the attitude that this divergence is
regulated so that the entropy in this thermal atmosphere
below some radius like $4M$ (the precise value of which
doesn't matter much, since the assumed entropy density
is this region is of the order of $\alpha/M^3$)
is the black hole entropy $S_{\rm bh} \approx A/4 = 4\pi M^2$.
Then one can say that if the shell
is put at a much larger radius,
the entropy of the thermal Hawking radiation
outside $4M$ or so would be matter entropy $S_{\rm m}$
that would add to $S_{\rm bh}$,
which is certainly an uncontroversial assumption.

	What I have found from my consideration
of having the shell squeezed in adiabatically
is that if the shell is put much nearer the horizon
than a radius of $4M$ or so is, then the entropy
is correspondingly less than the usual black hole entropy
$S_{\rm bh} \approx A/4 = 4\pi M^2$.
Because the thermal atmosphere is restricted
from filling up the region to $4M$ or so,
it does not have the entropy needed to make
the total entropy as large as $A/4$.

	Another way to justify this result
is to start with a zero-entropy perfectly reflecting shell
at $R = 4M$ ($W = 1/2$) and vacuum outside,
so that the initial entropy is roughly $A/4$
(i.e., up to an additive correction of the order
of unity, plus a possible correction of the order
of $\ln{A}$).
Then, without changing the total entropy,
construct a new zero-entropy
perfectly reflecting shell
at a value of $W = 2M/R$ much nearer unity.
Next, adiabatically pump out the thermal radiation
between the two shells, so that the region in between
becomes a vacuum region with zero entropy.
Finally, adiabatically discard the outer zero-entropy shell.
If the thermal radiation pumped out is discarded
(e.g., sent to infinity) and is no longer counted
in the entropy of the configuration,
and if discarding the outer shell does not change the entropy,
this whole process should reduce the entropy being counted by
that of the original thermal radiation
in the region between the shells.
This entropy is the difference between that of the thermal
radiation and that of the vacuum in that region,
so there should be no inherent ambiguities from
any supposed renormalization of the entropy.
If one ignores the backreaction of this radiation
on the metric, one can calculate the entropy
from the Hawking temperature and from the
radiation eigenmodes and their frequencies
in the region between the two shells.
When the inner shell has $1 - W \ll 1$,
this entropy is given, to a good approximation,
by the difference between the values of $S(M,W)$
for the two values of $W$ corresponding to the two shells.

\section{Fundamental Limitations on the Range of \\
Validity of the Black Hole Entropy Formula}

	The next question is the range of $W$
over which one would expect that Eq. (\ref{eq:49})
is approximately valid.
For very small $W$ or very large $R$, one essentially
has a black hole of mass $M$ surrounded by
a much bigger volume, $V \sim 4\pi R^3/3$,
of radiation in nearly flat
spacetime with Hawking temperature $1/(8\pi M)^{-1}$,
energy density roughly $3\alpha/(32\pi M^4)$,
and entropy density roughly $\alpha/M^3$.
The dominant term for
the total energy of the radiation is
$E_r \sim \alpha R^3/(8M^4)$,
and from Eq. (\ref{eq:49}), the dominant term for
the total entropy of the radiation is
$S_r \sim 4\pi\alpha R^3/(3M^3)$. 
This agrees with the standard expression
for the entropy of thermal radiation of energy $E_r$
in a volume $V$,
 \begin{equation}
 S_r = {4\over 3}(a_r V)^{1/4} E_r^{3/4}
 \sim {4\pi\over 3}\alpha^{1/4}(8 R E_r)^{3/4}.
 \label{eq:59}
 \end{equation}

	For fixed total energy $M_{\infty} = M + E_r \ll R$,
the total entropy
 \begin{equation}
 S \approx 4\pi M^2 + S_r
 \sim 4\pi(M_{\infty}-E_r)^2
  + {4\pi\over 3}\alpha^{1/4}(8 R E_r)^{3/4}
 \label{eq:60}
 \end{equation}
is indeed extremized for
 \begin{equation}
 E_r \sim {\alpha R^3 \over 8 M^4}
  = {\alpha R^3 \over 8(M_{\infty}-E_r)^4},
 \label{eq:61}
 \end{equation}
but the extremum is a local entropy maximum
if and only if $5E_r \leq M_{\infty}$ or $4E_r \leq M$
\cite{BHbox},
which implies that one needs $R \leq (2M^5/\alpha)^{1/3}$ or
 \begin{equation}
 W \geq \left({4\alpha\over M^2}\right)^{1/3}
 \label{eq:62}
 \end{equation}
for thermodynamic stability.

	For smaller values of $W$ (larger values of $R$),
the radiation energy $E_r$ is more than $20\%$ of the
total available energy $M_{\infty}$ (assumed to be held fixed),
and then if the black hole emits some extra radiation
and shrinks, it heats up more than the radiation does,
leading to an instability in which the black hole
radiates away completely. On the other hand,
if the black hole absorbs some extra radiation,
it will grow and cool down more than the surrounding radiation,
therefore cooling down more and absorbing more radiation,
until the radiation energy $E_r$ drops to the lower
positive root of Eq. (\ref{eq:61}),
which is less than $0.2 M_{\infty}$ and hence is
at least locally stable.
However, for larger values of $W$,
obeying the inequality (\ref{eq:62}),
the net feedback to extra emission or absorption
by the black hole is negative,
so that the corresponding configuration is locally stable
with fixed total energy $M_{\infty}$.

	At the opposite extreme,
the question is how small $1-W$ can be.
Here the fundamental limit is the Planck regime,
which is the boundary of the semiclassical
approximation being used in this paper.
The Boulware vacuum energy density $\rho_B$
just outside a massless shell
(so that the mass just outside, $M_{+}$,
is very nearly the same as the black hole mass $M$;
for positive shell mass $\mu$, $\rho_B$
would have an even greater magnitude)
is, for very small $U = 1-W$,
$\rho_B \sim - 3\alpha/(32\pi M^4 U^2)$.
Suppose the semiclassical theory is valid
until the orthonormal Einstein tensor component
$G_0^0 = -8\pi\rho_B \sim 3\alpha/(4 M^4 U^2)$
reaches a maximum value of, say, $C_M$,
which would be expected to be of order unity
(orthonormal curvature component of the order
of the Planck value).
This gives the restriction
 \begin{equation}
 U = 1-W \geq \left({3\alpha\over 4 C_M M^4}\right)^{1/2}.
 \label{eq:63}
 \end{equation}
 
	For $U = 1-W \ll 1$, the spatial distance
from the shell to the horizon is $D \sim 4M U^{1/2}$,
so this restriction on $U$ gives a minimum distance the
shell can be from the horizon:
 \begin{equation}
 D \geq \left({192\alpha\over C_M}\right)^{1/4},
 \label{eq:64}
 \end{equation}
in Planck units, as all quantities are in this paper
unless otherwise specified.

	If we combine the restriction (\ref{eq:63})
with the lower bound on $W$ from Eq. (\ref{eq:62})
and re-express the combined restriction as a restriction on
the radius $R$ of the shell, we get
 \begin{equation}
 2M+{1\over M}\sqrt{3\alpha\over C_M} \leq R
  \leq \left({2M^5\over\alpha}\right)^{1/3}.
 \label{eq:65}
 \end{equation}
Alternatively, in terms of the distance $D$
of the shell to the horizon (which is $D \sim R$
for $R \gg 2M$), we get
 \begin{equation}
 \left({192\alpha\over C_M}\right)^{1/4} \leq D
  \leq \left({2M^5\over\alpha}\right)^{1/3}.
 \label{eq:64b}
 \end{equation}

	If we now insert the restriction (\ref{eq:63})
or (\ref{eq:64}) into the asymptotic form of the total
entropy (\ref{eq:49}) for $U = 1-W \ll 1$, which is
 \begin{eqnarray}
 S(M,W) &\sim& 4\pi M^2 - {32\pi\alpha\over U}
 	\sim 4\pi M^2 \left( 1 - {128\alpha\over D^2} \right)
 \nonumber \\
	&\sim& 4\pi M_{\infty}^2 - {8\pi\alpha\over U}
	\sim 4\pi M_{\infty}^2 \left( 1 - {32\alpha\over D^2} \right),
 \label{eq:65b}
 \end{eqnarray}
we get the limitation
 \begin{equation}
 S(M,W) \geq 4\pi M^2 \left( 1 - 16 \sqrt{\alpha C_M \over 3} \right)
 = 4\pi M^2
  \left( 1 - \sqrt{C_M \over 135\pi}(n_b + {7\over 8}n_f) \right).
 \label{eq:66}
 \end{equation}
This can be re-expressed as a limitation on how much
the area $A$ of a black hole can exceed
four time the entropy, $4S$:
 \begin{equation}
 A - 4S \leq A\sqrt{C_M \over 135\pi}(n_b + {7\over 8}n_f).
 \label{eq:66b}
 \end{equation}

	Therefore, unless we have
$N \equiv n_b + 7n_f/8$, the effective number of
one-helicity particles, comparable to or greater than
$\sqrt{135\pi/C_M} \approx 20.6/\sqrt{C_M}$,
the fractional increase of the black hole area $A$
above $4S$ is restricted to be rather small,
though even just $N=4$ from two-helicity
gravitons and photons would give a fractional increase of
about 19\% if the curvature limitation $C_M$
is one in Planck units.

\section{Limitations from Imperfectly Reflecting Shells}

	Another limitation on the reduction of entropy below $A/4$
by a reflecting shell is the fact that no shell can be a perfect
reflector.  For example, Smolin has argued
\cite{Smo}
that no realistic shell can be a good reflector
of gravitational radiation.
Therefore, the entropy of the gravitational radiation part of
the black hole thermal atmosphere cannot be significantly
reduced by surrounding the hole with a realistic shell.

	Strictly speaking, no shell is a perfect reflector
of any radiation, so if one waits for a sufficiently long
time that true thermal equilibrium of the radiation sets in,
no shell can stop the region outside from also being thermal.
In fact, if one waits long enough for the shell itself to come into
complete thermal equilibrium with the radiation, the shell
will evaporate and become part of the radiation,
with, for example, most of its baryons either
decaying, falling into the hole, or getting expelled.
However, one can consider squeezing the black hole atmosphere with
a shell over a shorter timescale than the timescale
for the shell to disintegrate.
If the shell is a very good (but not perfect)
reflector of some kinds of radiation
(apparently never possible for gravitational radiation),
then one can imagine squeezing the shell sufficiently slowly
that it is nearly adiabatic
but sufficiently rapidly that the squeezing is
over a timescale short in comparison
with the timescale for a significant amount of radiation
to leak through the shell.
Then the entropy of that kind of radiation can be greatly
reduced from its thermal values outside the shell,
and hence the total entropy can be reduced by the reduction
of the entropy in the thermal atmosphere of the kind
of radiation that is practically completely confined
to lie within the shell for that intermediate timescale.

\section{Further Practical Limits on Entropy Reduction Below $A/4$}

	We found in Eq. (\ref{eq:49})
that for a neutral spherical black hole
of area $A = 4\pi M^2$ surrounded by
a perfectly reflecting shell at $R-2M \ll M$,
the entropy is roughly
 \begin{equation}
 S \approx {1\over 4}A - {32\pi\alpha R\over R-2M}
   = {1\over 4}A - {n_b + {7 \over 8} n_f \over 360 (1-2M/R)}.
 \label{eq:90}
 \end{equation}
The last term represents the leading term
for the reduction of the entropy below one-quarter the area.
Let us ask how large this term can be
for various assumptions about the shell.

	First, consider the case that the shell
is held up entirely by its own stresses, with
no external forces (other than gravity) on it.
In particular, we shall consider the static shell
junction conditions (\ref{eq:37}) and (\ref{eq:38}),
applying the strong energy condition to the shell
so that its surface stress obeys the inequality
$S_2^2 = S_3^3 \leq -S_0^0$.  As we shall soon see,
it then turns out that $U_{-} = 1-2M_-/R \approx 1-W = 1-2M/R$
cannot be very small, so the terms involving the
pressures inside and outside the shell are then negligible.
Then the strong energy condition applied to the
junction conditions (\ref{eq:37}) and (\ref{eq:38})
imply that $U_{+}U_{-} \geq 1/25$, and since Eq. (\ref{eq:37})
implies that a shell with positive local mass has
$U_{+} < U_{-}$, we see that $1-W \approx U_{-} > 1/5$,
or $R > 2.5M$.  If Eq. (\ref{eq:90}) applied for such
a large value of $1-W$, it would then give
 \begin{equation}
 {1\over 4}A - S \approx  {n_b + {7 \over 8} n_f \over 360 (1-2M/R)}
 < {n_b + {7 \over 8} n_f \over 72},
 \label{eq:91}
 \end{equation}
a quite negligible decrease in the entropy,
unless somehow $n_b + (7/8) n_f$ is very large.
This decrease could indeed be dominated
by effects that I have ignored,
such as Casimir-energy effects,
and, to an even greater degree,
by possible terms proportional to $\ln{M}$.

	Next, consider the case that the shell
has charge $Q$, so that its electrostatic repulsion holds it up.
Since we found above that the stresses within the surface
of the shell are quite ineffectual in holding up the shell
at $R-2M \ll M$, let us drop them from the junctions equations
but add the tension of the electromagnetic field
outside the shell and assume that that tension is much greater
than the radial pressures (or tensions) of the quantum fields.
Then the junction conditions (\ref{eq:37}) and (\ref{eq:38})
become
 \begin{equation}
 {\mu\over R} = U_{-}^{1/2} - U_{+}^{1/2}
 \label{eq:92}
 \end{equation}
and
 \begin{equation}
 0 = 8F = {\mu\over R} + U_{-}^{-1/2}
  - (1 - {Q^2\over R^2})U_{+}^{-1/2},
 \label{eq:93}
 \end{equation}.

	Now for fixed charge-to-mass ratio $Q/\mu$,
if we let $\gamma = (\mu/R)/U_{-}^{1/2} < 1$,
Eq. (\ref{eq:92}) implies that
$U_{+}^{1/2} = (1-\gamma)(\mu/R)$,
which when inserted into Eq. (\ref{eq:93}) gives
 \begin{equation}
 {1\over 1-2M/R} \approx {1\over U_{-}}
  = 1-\gamma+\gamma(Q/\mu)^2 < (Q/\mu)^2.
 \label{eq:94}
 \end{equation}
If we take the charge-to-mass ratio of an electron,
we get $(Q/\mu)^2 \approx 4.17\times 10^{42}$.
If we then suppose that somehow a shell of electrons
reflects electromagnetic (but not other) radiation
and thereby manages to keep the electromagnetic field
in its Boulware vacuum state outside the shell
(rather than in the Hartle-Hawking thermal state
that exists within the shell), then $n_b = 2$
(from the two helicities of photons) and $n_f=0$,
so one gets
 \begin{equation}
 {1\over 4}A - S \approx  {1 \over 180 (1-2M/R)}
 < 2.31\times 10^{40}.
 \label{eq:95}
 \end{equation}
 
 	Of course, there are severe problems in attaining
anything near this limit.  First, electrons in a shell
around a black hole, even if in static equilibrium
as I have calculated they can be, will not be in
stable equilibrium, and some unknown mechanism
would have to be invoked to keep the shell in place.
Second, without specifying how the electrons are to be kept in place,
it is hard to say how they will respond to the black hole thermal
radiation impinging upon them from below.
However, it is interesting that the upper limit
given by Eq. (\ref{eq:95}) for the reduction in the entropy
below one-quarter the (neutral) black-hole area,
from a shell held up by electrostatic forces,
is so large (because the charge-to-mass ratio
of an electron is so large).

	For a somewhat more nearly realistic
example of a shell around a black hole,
consider a thin aluminum foil that is charged
so that, like the shell of pure electrons,
the electrostatic forces balance the gravitational forces.
In this case there will be limitations from
the mass density $\rho$ of the foil,
the minimum practical thickness $\tau$ of the foil
and the maximum charge per surface area, $\sigma$,
that it can hold.

	In the earlier and longer version of this paper
\cite{BHentropy},
I took the aluminum foil to have the parameters
(in conventional and Planck units respectively)
 \begin{equation}
 \rho \approx 2.70 \:{\rm g}/{\rm cm}^2
  \approx 5.23\times 10^{-94},
 \label{eq:103}
 \end{equation}
 \begin{equation}
 \tau = 0.0005 \:{\rm cm} \approx 3.09\times 10^{29}
 \label{eq:105}
 \end{equation}
(about 100 times the Meissner magnetic penetration depth
of about 50 nm
\cite{Mei},
so that the shell should be a very nearly perfect
reflector of the electromagnetic part of
the Hawking radiation inside it),
and
 \begin{equation}
 \sigma \approx 3.34\times 10^{12} \: e/{\rm cm}^2
 \approx 7.46\times 10^{-55}
 \label{eq:110}
 \end{equation}
(so that if the electric surface charge density
$\sigma$ were an excess of electrons,
the probability for one to tunnel off
would be a very small number, chosen to be
$\exp{(-100)}$, in some suitable atomic time unit).
I then calculated that the
charge-to-mass ratio of the aluminum foil is
 \begin{equation}
 {Q\over\mu}={\sigma\over\rho\tau}
 \approx 4.61\times 10^9.
 \label{eq:112}
 \end{equation}

	Then, by the same analysis used above for
the pure electron shell, one finds that if one takes
$\mu/R = U_{-}^{1/2} = \mu/Q$, one gets
 \begin{equation}
 {1\over 1-2M/R} = \left({Q\over\mu}\right)^2
 \approx 2.12\times 10^{19},
 \label{eq:113}
 \end{equation}
and hence the reduction of the entropy from excluding
the thermal photons from above the shell is
 \begin{equation}
 \Delta S \equiv {1\over 4}A - S \approx {1 \over 180 (1-2M/R)}
 \approx 1.18\times 10^{17}.
 \label{eq:114}
 \end{equation}

	This is extremely tiny in comparison with
the total entropy of the black hole,
 \begin{equation}
 S \approx 4\pi M^2 \approx {1\over 4\pi\sigma^2}
  \approx 3.58\times 10^{106},
 \label{eq:118}
 \end{equation}
but it is very large in absolute value.
Indeed, it is much larger than any
correction to the entropy that I have ignored
if, as I would expect, such corrections
are no larger than some number of the order of
$\ln{M}$, which is in this case about 121.4.

This reduction in the entropy from the naive
value of $A/4$ (or of this plus or minus
a correction of the order of $\ln{M}$)
means that the number of states is fewer by a factor
of about
 \begin{equation}
 e^{\Delta S} \sim 10^{51\,000\,000\,000\,000\,000},
 \label{eq:115}
 \end{equation}
which is quite a large factor.
Therefore, even though the reduction in the entropy
is a tiny fraction of the total entropy,
it is large in absolute value, and hence,
when exponentiated, it gives an enormous factor
in the reduction of the total number of states.
It is this sense in which the entropy
of a black hole can be significantly reduced below
$A/4$ by restricting its thermal atmosphere
to lie below that of a reflecting shell
at very small values of $1 - 2M/R$.

\newpage
\section{Conclusions and Acknowledgments}

	Thus we have seen that by placing a reflecting shell
around a black hole, we can make the entropy have a value
that is below one-quarter its area
(or one-quarter its area plus a positive or negative correction
of the order of the logarithm of the area in Planck units).
If we are allowed an idealized perfectly reflecting shell
that can be placed within roughly one Planck length of the horizon,
then this entropy reduction can be of the same order as
the area of the hole.  For a more realistic shell,
such as a superconducting aluminum foil,
the entropy reduction can only be a tiny fraction of the area,
but it still can be huge in absolute units
(much larger than other corrections arising from,
say, logarithms of the black hole mass or area),
markedly reducing the number of black hole states from
what would be erroneously estimated by exponentiating
one quarter the horizon area.

	I have benefited from conversations with,
among many others whose names did not immediately come to mind,
Valeri Frolov, Frank Hegmann, Akio Hosoya, Satoshi Iso, Sang-Pyo Kim,
Frank Marsiglio, Sharon Morsink, Shinji Mukohyama, Jonathan Oppenheim,
Lee Smolin, L. Sriramkumar, Bill Unruh, and Andrei Zelnikov.
Comments of the referees of
\cite{BHentropy}
have led me to try to explain more precisely what I have calculated
and to give additional justifications for
my resulting black hole entropy formula (\ref{eq:49}).
Part of this work was done at the Tokyo Institute
of Technology under the hospitality of Akio Hosoya,
and part was done at the Haiti Children's Home
of Mirebalais, Haiti, under the hospitality
of Patricia and Melinda Smith,
while adopting Marie Patricia Grace Page.
This research was supported in part by
the Natural Sciences and Engineering Research
Council of Canada.

\end{document}